\documentclass{article}
\usepackage{amsmath,amsfonts}
\usepackage{graphicx,psfrag,epsf}
\usepackage{enumerate}
\usepackage{natbib}
\usepackage{color}
\usepackage[a4paper]{geometry}
\newcommand*{\Scale}[2][4]{\scalebox{#1}{$#2$}}%
\begin{document}
\title{Optimal Design of Experiments for Nonlinear Response Surface Models}
\author{Yuanzhi Huang \\
    School of Mathematics and Statistics\\
    University College Dublin, Dublin, Ireland\\
    Insight Centre for Data Analytics\\
    University College Dublin, Dublin, Ireland\\
    (Yuanzhi.Huang@ucd.ie)\\
    and \\
    Steven G.\ Gilmour \\
    Department of Mathematics\\
	King's College London, Strand, London, United Kingdom\\
    and \\
    Kalliopi Mylona \\
    Department of Statistics\\
    Universidad Carlos III de Madrid, Madrid, Spain\\
    Department of Mathematics\\
	King's College London, Strand, London, United Kingdom\\
    and \\
    Peter Goos \\
    Department of Biosystems\\
    KU Leuven, Leuven, Belgium\\
	Department of Engineering Management\\
	University of Antwerp, Antwerp, Belgium}
\maketitle

\bibliographystyle{apalike} 

\begin{abstract} 
Many chemical and biological experiments involve multiple treatment factors and often it is convenient to fit a nonlinear model in these factors. This nonlinear model can be mechanistic, empirical or a hybrid of the two. Motivated by experiments in chemical engineering, we focus on D-optimal design for multifactor nonlinear response surfaces in general. In order to find and study optimal designs, we first implement conventional point and coordinate exchange algorithms. Next, we develop a novel multiphase optimisation method to construct D-optimal designs with improved properties. The benefits of this method are demonstrated by application to two experiments involving nonlinear regression models. The designs obtained are shown to be considerably more informative than designs obtained using traditional design optimality algorithms.
\end{abstract}   


\section{Nonlinear Multifactor Models}\label{sec:intro}
In chemical and biological studies, experimenters often wish to explore mechanisms relating controllable input variables (treatment factors) to observed outputs. After collecting data, they fit a statistical model describing the relationship between the levels of the treatment factors and the experimental responses. In this paper, we are interested in the estimation of the model parameters, in experiments in which a model is known or assumed at the design stage. A statistical model can be either mechanistic (theoretical) or empirical and illustrative examples of each are given in Section \ref{sec:applications}. Unlike common empirical models, mechanistic models tend to be nonlinear but more frugal in the use of parameters. For some discussion of mechanistic studies, see \citet{bo+6}. The mechanistic model is often derived from scientific theories and formulas, though it is generally still an approximation to the true response function. It often allows more scientifically informative interpretations than some empirical models. 

Although less often used, these benefits apply even when the unknown mechanism involves multiple treatment factors. In the event that there are multiple input factors, however, deriving a full mechanistic model can be very hard and therefore no such model might be available at the start of an experiment. Under these circumstances, it is advisable to resort to empirical modelling techniques, while still taking into account the known parts of the mechanism when building the model. These kinds of models can be referred to as \textit{hybrid models}, incorporating both mechanistic and empirical features. Particularly for mechanistic models, it is recommended to use model-oriented optimal design of experiments, since no standard designs exist for such models. This approach requires advanced optimisation methods \citep{ch, got, gi, go+1, ow}. Little work has been done on designing multifactor experiments for hybrid models or multifactor nonlinear models in general. This is the focus of this paper. Optimal design for a few specific classes of nonlinear models have been studied in detail. In particular, some early work exists on two-factor inverse polynomial response surface models for agricultural experiments \citep{mea}, as well as several studies with generalised linear models, e.g.\ \citet{woo,stu}, including discrete choice models \citep{rgf} and recent pharmacokinetic studies with nonlinear mixed models \citep{bog+1}.

In Section \ref{sec:applications}, two experiments of the type which motivated the work presented are introduced. In Section \ref{sec:review}, commonly used existing methods for constructing optimal designs are described. These methods are then built on in Section \ref{sec:continuous} to describe a novel continuous optimisation algorithm for finding optimal designs. The application of the new algorithm to the illustrative applications is described in Section \ref{sec:results}. In Section \ref{sec:multiphase}, our continuous optimisation method is combined with traditional ones, to obtain a computationally efficient optimisation which produces the best known designs for these problems. This combined approach is applied to the two motivating applications in Section \ref{sec:moreresults}. Final comments and recommendations are made in Section \ref{sec:discussion}.

\section{Illustrative Applications}\label{sec:applications}
In order to see the need for improved design, we use two experiments which were actually run using central composite designs (CCDs) and are typical of the types of experiment performed in chemical engineering studies. In both cases, multifactor nonlinear models were useful for describing the data, though the experiments were not designed to be optimal for these models. The CCDs actually used give a benchmark against which improved designs can be compared.

\subsection{Example 1: Multifactor Mechanistic Model}
Our first motivating example is described in \citet[chapters 11-12]{bo}, where both an empirical full second-order linear model and a mechanistic nonlinear model were fitted to the same data. A continuous stirred reactor was operated, for which the chemists wished to approximate the response surface and interpret the mechanism it may imply. The assumed mechanism, which was derived after the experiment was conducted, describes a two-step consecutive decomposition reaction of a chemical solution. The yield $\eta$ (\%) is the amount of the desirable product that has been formed, which is the primary response. The experiment involved three input factors: the rate $R$ (L/min) at which the chemical flows into the reactor, the catalyst concentration $C$ and the temperature ($^{\circ}$C). Yield data were obtained using a 24-run spherical CCD. 

The full second-order linear model involving 10 parameters was fitted using the scaled variables
$$x_1=\frac{\mathrm{log}(R)-\mathrm{log}(3)}{\mathrm{log}(2)}, x_2=\frac{\mathrm{log}(C)-\mathrm{log}(2)}{\mathrm{log}(2)}, x_3=\frac{T-80}{10}.$$
The characteristics of this chemical reaction can be better understood if we fit the mechanistic model 
\begin{equation}
\eta=\frac{C^{\theta_1}\theta_0 R \mathrm{exp}(\theta_2 X)}{(R + C^{\theta'_1}\theta'_0\mathrm{exp}(\theta'_2 X))(R + C^{\theta_1}\theta_0\mathrm{exp}(\theta_2 X))}+\varepsilon,
\label{eq1}
\end{equation}
where $X = 0.0028344-1/(T+273)$ and $\varepsilon$ is the error term. This model is derived from a number of differential equations and the Arrhenius equation for temperature effects. In this paper we will show how to plan a new experiment in order to precisely estimate the parameters $\theta_0$, $\theta'_0$, $\theta_1$, $\theta'_1$, $\theta_2$ and $\theta'_2$, of the mechanistic model in Model (\ref{eq1}).

\subsection{Example 2: Hybrid Nonlinear Model}
In the second motivating example, the experimenters' interest is in the enzymatic depolymerisation mechanism of a dextran substrate \citep{mo}. In a stirred-cell membrane reactor, endodextranase is used as the enzyme activator, while the reactants are different kinds of oligodextrans. The treatment factors are the substrate concentration $S$ (2.5-7.5\% in weight/volume), the enzyme concentration $E$ (0.625-62.5 Units $\mathrm{ml}^{-1} \times S$) and the transmembrane pressure $P$ (200-400 kPa). The response we consider is the substrate conversion rate $\xi$ (\%). The data were obtained using an 18-run face-centred CCD.

No mechanistic model could be found to describe the response surface for $\xi$, so the alternative was to use an empirical approximation. In spite of the simplicity of deriving purely empirical polynomial models, however, it is worth considering a hybrid model which could have some mechanistic implications and so catalyse scientific reasoning. First, we can observe a negative effect from the experimental data: the higher the substrate concentration, the lower the conversion rate $\xi$. This suggests a smooth function of the form $\mathbb{E}(\xi/(100-\xi))=\gamma_1'S/(\gamma_2'+S)$, where $\mathbb{E}(\xi/(100-\xi))$ is the expectation of the transformed response and $\gamma_1'$ and $\gamma_2'$ are constants with $\gamma_1'\geq 0$ and $\gamma_2' \in (-2.5,0)$. The parameter $\gamma'_1$ is then replaced with an empirical exponential function of the coded variables $x_{\Scale[0.5]{\mathrm{E}}}=\mathrm{log}_{10} (E/6.25) \in [-1,1]$ and $x_{\Scale[0.5]{\mathrm{P}}}=(P-300)/100 \in [-1,1]$. This results in the hybrid nonlinear model given by
\begin{align}
\frac{\xi}{100-\xi}=\frac{\mathrm{exp}(a_0+a_1x_{\Scale[0.5]{\mathrm{E}}}+a_2x_{\Scale[0.5]{\mathrm{P}}}+a_3x_{\Scale[0.5]{\mathrm{E}}}^2+a_4x_{\Scale[0.5]{\mathrm{P}}}^2)S}{a_5+S}+\varepsilon,
\label{eq2}
\end{align}
where $\varepsilon$ is the error term and $a_0$, $a_1$, $a_2$, $a_3$, $a_4$ and $a_5$ are the six model parameters. For details of the model derivation from our conjectures, see \textbf{APPENDIX \ref{sec:A}}. In this paper, we show how to create optimal experimental designs for the model in Model (\ref{eq2}).

\section{Methods for Optimal Design}\label{sec:review}
\subsection{Local D-Criterion for Nonlinear Models}
To obtain an exact experimental design $\mathit{X}$, we need to choose a level for each of the $v$ treatment factors for each of the $n$ runs. For $i=1,2,\ldots,n$, the $i$th response is $Y_i=f(\mathit{X}_i,\boldsymbol{\theta})+\varepsilon_i$, where $\mathit{X}_i$ is the $i$th row of the $n \times v$ design matrix $\mathit{X}$ and $\boldsymbol \theta$ denotes the set of $p$ parameters. Under ordinary least squares estimation, the errors $\varepsilon_i$ must be uncorrelated and come from identical distributions with zero mean and constant variance $\mathbb{V}(\varepsilon_i)=\sigma^2$. 

If the function $f(\mathit{X}_i,\boldsymbol{\theta})$ is linear in the parameters, the matrix form of the model is $\mathbf{Y}=\mathit{F}\boldsymbol \theta+\boldsymbol \varepsilon$. Here, $\mathbf{Y}$ is the column vector of the $n$ responses, $\boldsymbol \varepsilon$ is the column vector of the $n$ error terms, and $\mathit{F}$ is the $n \times p$ model matrix based on the design $\mathit{X}$. 

We assume the primary purpose of the experiment to be the precise estimation of the parameters of the model. This requires us to minimise the elements of the $p \times p$ variance-covariance matrix $\mathbb{V}(\hat{\boldsymbol \theta}) = (\mathit{F}^\mathrm{T}\mathit{F})^{-1} \sigma^2$, where we assume, without loss of generality, that $\sigma^2 = 1$ when constructing designs and $\mathit{F}^\mathrm{T}\mathit{F}$ is the Fisher information matrix. To ensure that $\mathit{X}$ contains the maximal amount of information about $\boldsymbol \theta$, we can search for a design $\mathit{X}$ that maximises the D-criterion function $\phi=\mathrm{log}\left|\mathit{F}^\mathrm{T}\mathit{F}\right| \in (-\infty, +\infty)$. If $\phi_A$ and $\phi_B$ are the D-criterion function values of any two designs $\mathit{X}_\mathrm{A}$ and $\mathit{X}_\mathrm{B}$ respectively, the relative efficiency of $\mathit{X}_\mathrm{A}$ with respect to $\mathit{X}_\mathrm{B}$ is 
$$\mathit{eff}=\frac{\mathrm{exp}(\phi_A/p)}{\mathrm{exp}(\phi_B/p)}100\% \in (0,+\infty).$$
To derive the D-criterion function of any nonlinear model (regardless of its functional form), the most widely used approach is to linearise $f(\mathit{X}_i,\boldsymbol{\theta})$ in $\boldsymbol{\theta}$ (see e.g.\ \citet[chapter 17]{at}) by means of a first-order expansion about a selected centre $\boldsymbol{\theta}^0$, giving
\begin{equation}
f(\mathit{X}_i,\boldsymbol{\theta}) \approx f(\mathit{X}_i,\boldsymbol{\theta}^0)+\sum_{j=1}^p \left(\frac{\partial f(\mathit{X}_i,\boldsymbol \theta)}{\partial \theta_j} \Bigl | _{\theta_j=\theta^0_j}\right)(\theta_j-\theta^0_j),
\label{eq3}
\end{equation}
where $\theta_1, \ldots, \theta_p$ are the elements of $\boldsymbol{\theta}$. The $i$th row of $\mathit{F}$ is then
$$\mathit{F}_i=\frac{\partial f(\mathit{X}_i,\boldsymbol \theta)}{\partial \boldsymbol{\theta}} \Bigl |_{\boldsymbol{\theta}= \boldsymbol{\theta}^0}$$
and the corresponding information matrix is given by $\mathit{F}^\mathrm{T}\mathit{F}$. For a nonlinear model, $\phi$ is called the ``local'' D-criterion function, since its value is associated with the substitution $\boldsymbol{\theta}=\boldsymbol{\theta}^0$. The choice of the prior $\boldsymbol{\theta}^0$ is generally based on similar experiments reported in the literature, the experimenters' expertise or even a ``best guess'' about the response surface. The design optimises $\phi$ when $\boldsymbol{\theta} = \boldsymbol{\theta}^0$ and usually is optimal or close to optimal when $|\boldsymbol{\theta}-\boldsymbol{\theta}^0|$ is small. As the difference in $|\boldsymbol{\theta}-\boldsymbol{\theta}^0|$ increases, the design might be suboptimal, though it is still valid in the sense that it gives asymptotically unbiased estimators of all the parameters. 

Because of the dependence of the Fisher information on the model, the parameter prior, and the number of experimental runs, the optimal designs for nonlinear models may be quite different from those for second-order or even higher-order linear models. Hence we need to find the optimal design for the proposed nonlinear model if feasible. To search for an exact design $\mathit{X}$ comprising $n \times v$ coordinates or factor settings, we develop an algorithm that can derive and quickly integrate the local D-criterion function in an iterative search and optimisation. The algorithm should be able to work with any nonlinear parametric model, independent of model assumptions, and, up to computational limitations, any experimental size. The essential inputs to our algorithm are the model function $f(\mathit{X}_i,\boldsymbol{\theta})$, the number of experimental runs $n$, the $v$-dimensional experimental region $\mathcal{X}$ and initial values of the parameters $\boldsymbol{\theta}_0$. 

\subsection{Iterative Discrete Optimization}
\citet{fe} built the theoretical framework for the earliest algorithm for locally D-optimal design (LDOD). The algorithm requires a discretisation of the continuous design region $\mathcal{X}$, defined by the possible treatment factor levels, to obtain a set $\boldsymbol \Omega$ of $N$ candidate points. For instance, for a first-order linear model with three factors, two candidate levels should be defined for each factor (i.e.\ the maximum and minimum). In that case, there are $2 \times 2 \times 2 =8$ candidate points in total. A full second-order linear model demands at least one more candidate level for each factor, such that there would typically be $3 \times 3 \times 3 =27$ candidate points. Defining candidates is much harder for nonlinear models, since it is generally unknown what will be good candidate points. Having defined the candidate set $\boldsymbol \Omega$, we then apply \textit{discrete optimisation} through numerous iterations, each of which can exchange at most one point of the design $\mathit{X}$ with one candidate point. The more candidate points are included, the more reliable the optimisation is, but the higher the computational time. 

Despite the dependence between iterations, $\mathit{X}$ can usually be globally optimised, though there is no guarantee of this. To speed up the computation, the modified Fedorov exchange algorithm \citep{coo} streamlines the iterative procedure of the Fedorov exchange algorithm, i.e.\ up to $n$ points of $\mathit{X}$ can be updated in each iteration and an exchange is executed whenever a clear improvement in the criterion function is achievable. A point exchange algorithm (PEA) we adapted for nonlinear models which, like all other algorithms we present in this paper, is implemented in the Matlab environment is as follows:

\begin{enumerate} 
  \item[1] Generate an initial nonsingular $n$-run design $\mathit{X}$ by means of $n$ independent random draws (with replacement) from the candidate set $\boldsymbol \Omega$. After substituting $\boldsymbol{\theta}=\boldsymbol{\theta}^0$, compute $\mathit{F}$ and $\phi$.  
  \item[2] Set index $i$ to 1 and $\Upsilon$ to 0. 
  \item[3] Evaluate the relative improvement function $d_i$ for each possible point exchange between the $i$th row of $\mathit{X}$ and one of the candidate points in $\boldsymbol \Omega$. 
  \item[4] If the maximal $d_i$ exceeds the critical value, execute the corresponding substitution $\mathit{X}_i=\mathit{X}_{\mathrm{new},i}$ in $\mathit{X}$. Update $\mathit{F}$ and $\phi$ and set $\Upsilon$ to 1.
  \item[5] If $i < n$, set $i$ to $i+1$ and return to \textbf{STEP 3}.
  \item[6] If $\Upsilon=1$, return to \textbf{STEP 2}. Otherwise, save the current $\mathit{X}$ and the maximal local D-criterion value $\phi$.
  \item[7] Repeat \textbf{STEPS 1-6} for $\tau$ independent tries. 
  \item[8] Report the best of the designs found. That is the LDOD.
\end{enumerate}

In the above algorithm, we use the indicator variable $\Upsilon$, which takes the value 1 if at least one improvement has been made to $\mathit{X}$ in the current iteration of the algorithm. In that case, a complete new iteration will be performed after finishing the current one. We also define the relative improvement function $d_i=|\mathit{F}^\mathrm{T}\mathit{F}|_\mathrm{new}/|\mathit{F}^\mathrm{T}\mathit{F}|$, for $i=1,2,\ldots,n$, where $|\mathit{F}^\mathrm{T}\mathit{F}|_\mathrm{new}$ denotes the determinant of the updated information matrix after substituting $\mathit{X}_i=\mathit{X}_{\mathrm{new},i}$ in $\mathit{X}$. The iteration can take advantage of the updating function of $|\mathit{F}^\mathrm{T}\mathit{F}|$  for the D-criterion \citep{fe}. Sequentially maximising $d_i$ approximates maximising the D-criterion function $\phi$ defined in terms of $\mathit{X}$. A smaller critical value in \textbf{STEP 4} would lead to more iterations and updates of $\mathit{X}$. As a result, the design $\mathit{X}$ attained in \textbf{STEP 6} may converge further towards (and get stuck in) a local optimum. We discuss in Section \ref{sec:continuous} how to determine the critical value.

In a PEA, each iteration is broken into $n$ dependent steps, one for each row of $\mathit{X}$. Therefore, we should do a number of tries, i.e.\ run the algorithm for several random starting designs, to secure an efficient solution $\mathit{X}_\mathrm{d}$.

An alternative to a PEA is the coordinate exchange algorithm (CEA) of \cite{me}. For discrete optimisation, this uses a unidimensional subset $\boldsymbol \Omega_k$ at each step of the iteration, which consists of the candidate coordinate levels of the $k$th factor. Compared with the outline above, in \textbf{STEP 3}, instead of updating the current row $\mathit{X}_i$ in one step of the iteration, at most one coordinate $X_{ik}$ shall be updated. Specifically, $X_{ik}$ is replaced by the best candidate coordinate from $\boldsymbol \Omega_k$, for $k=1,2,\ldots,v$. In the CEA, \textbf{STEPS 3} and \textbf{4} of the PEA are replaced by 

\begin{itemize}
\item[3a] Set index $k$ to 1.
\item[3b] The relative improvement function $d_i$ is evaluated for each possible coordinate exchange between the $k$th coordinate in the $i$th row of $\mathit{X}$ and one of the candidate coordinates in $\boldsymbol \Omega_k$. 
  \item[4a] If the maximal $d_i$ value exceeds the critical value, execute the corresponding substitution $\mathit{X}_{ik}=\mathit{X}_{\mathrm{new},ik}$ in $\mathit{X}$, which maximises $d_i$. Update $\mathit{F}$ and $\phi$ and set $\Upsilon$ to 1.
\item[4b] If $k<v$, set $k$ to $k+1$ and return to \textbf{STEP 3b}.
\end{itemize}

Each iteration of the CEA takes $vn$ steps to separately optimise all the coordinates of $\mathit{X}$. When the experiment involves many runs and factors, this approach reduces the computational time compared with the PEA. On the other hand, the local search for optimal coordinates is less extensive than that of the PEA, so that fewer designs are evaluated in each iteration.

For solving very complicated and high-dimensional LDOD problems, some advanced heuristic search methods, such as particle swarm optimisation \citep{wow,ph}, have been used in the recent literature. These methods are able to learn from stochastic search and iteration, and tend to run faster than the PEA or the CEA, but often give suboptimal solutions \citep{blu,so,rgf} and can be very inefficient when users are unlucky with the choice of tuning constants. In low-dimensional cases, searching for LDODs is not too computationally expensive and we should use the PEA or the CEA for reliable solutions.

\section{A Novel Continuous Optimisation Method}\label{sec:continuous}
In this section, we present a new, improved, computing method which overcomes the main weaknesses of the PEA and CEA discussed in Section \ref{sec:review}, namely that the candidate points are limited to a fixed set and must be predefined. These weaknesses can lead to lower efficiencies, especially for nonlinear multifactor experiments. It is often hard to decide on suitable discrete candidates for nonlinear models and it is generally useful to explore points outside any finite set $\boldsymbol \Omega$. One could use a denser $\boldsymbol \Omega$, with smaller meshes and narrower space between adjacent candidate points within $\mathcal{X}$, but this would slow down the computation, and is not a fundamental solution to the problems of discrete optimisation. 

For most experiments using linear models with sufficient numbers of runs, the traditional discrete optimisation methods work well because equally spaced levels are generally either optimal or very close to optimal. In contrast, when the model is nonlinear, the optimal factor levels are generally not equally spaced and it is difficult in advance to propose a good candidate set. Hence, a \textit{continuous optimisation} of the relative improvement function $d_i$ over $\mathcal{X}$ in each iteration is more attractive. Below is a succinct outline of the \textit{continuous optimisation point exchange algorithm}:

\begin{enumerate}
  \item[] \textbf{STEPS 1-2} as in the PEA, but points of $\mathit{X}$ are drawn from $\mathcal{X}$ instead of $\boldsymbol \Omega$. 
  \item[3] Using the Nelder-Mead (or quasi-Newton) method, find the point in the design region $\mathcal{X}$ which numerically maximises $d_i$.
  \item[] \textbf{STEPS 4-8} as in the PEA.
\end{enumerate}

The main innovation in \textbf{STEP 3} is to use a continuous optimisation search over the entire design region, rather than just a discrete set of candidate points, as is done in the traditional PEA. A second novel aspect is that algebraic computing is used to derive the updating formula for $d_i$, based on the conventional updating formulae of \citet{fe}. An alternative version of the new algorithm based on coordinate exchange can be constructed in a similar way, the difference being that the optimisation is then done one coordinate at a time in \textbf{STEP 3} and $\mathit{X}$ is updated more often in each iteration. 

While the critical value in \textbf{STEP 4} has little impact in the discrete optimisation over a small number of candidates, it can advantageously be set to be smaller in the continuous optimisation so as to allow for minor iterative improvements on $\mathit{X}$ (i.e.\ a larger number of iterations on average). In both of our examples, 1.0001 is a reasonable critical value for the continuous optimisation, which allows for 4-6 iterations on average. 

We choose the Nelder-Mead method \citep{ne, pr} to maximise the function $d_i \in (0, +\infty)$ and thus the D-criterion function $\phi$ at each exchange. This maximisation is subject to the boundary constraint $\mathit{X}_i \in \mathcal{X}$. The Nelder-Mead simplex method excels at optimising complex functions \citep{pr}. Because the continuous optimisation is still based on iterative search over the design region $\mathcal{X}$, the solutions we find are guaranteed to be locally optimal only. In comparison, the optimal design algorithm in \citet{got} is based on a traditional CEA combined with the less accurate nonlinear optimisation method of Brent (1973, chapter 5), which has been implemented in the commercial software package JMP. Brent's method is for one-dimensional optimisation of coordinates in X and thus infeasible for the PEA. In revisiting our first application in Section \ref{sec:moreresults}, we will compare our algorithm with the CEA in combination with Brent's method.

For problems in which the total number of coordinates in design $\mathit{X}$ is small (e.g.\ $vn<10$), \citet{ch} introduced a method for direct optimisation of the entire design $\mathit{X}$. However, this approach is infeasible for large values of $vn$. Our continuous optimisation PEA and CEA can be considered to be a middle ground between this method and the traditional PEA/CEA.

\section{Numerical Results and Comparisons}\label{sec:results}
We now demonstrate how experiments can be optimally designed for nonlinear models by applying our improved PEA and CEA to the illustrative experiments in Section \ref{sec:applications}. To this end, we compare LDODs constructed using different approaches. 

\subsection{Example 1: Multifactor Mechanistic Model}
For the experiment from \citet{bo}, we compare the results from our continuous PEA and CEA to those of the traditional algorithms that use discrete optimisation. The benchmark we use for the LDOD is a face-centred CCD with two replicates of the axial points and four centre points. This is a modification of the design actually used to fit a cubic region of experimentation. The design region is taken to be $\mathcal{X}=[1.5,6] \times [1,4] \times [70,90]$. In \textbf{APPENDIX \ref{sec:B}}, additionally, we evaluate two alternative standard designs: 1) a spherical CCD with radius $\sqrt{2}$, which is similar to the design used but restricted to the narrower design region; 2) a Box-Behnken design. These three types of standard designs are all commonly used and require fewer runs than a full $3 \times 3$ factorial. To evaluate the local D-criterion function for Model \eqref{eq1}, the prior $\boldsymbol \theta^0$ is taken to be the nonlinear least squares estimate $\tilde{\boldsymbol \theta}=\{ \tilde \theta_0=5.90,~\tilde \theta'_0=1.15,~\tilde \theta_1=0.53,~\tilde \theta'_1=-0.01,~\tilde \theta_2=15475, ~\tilde \theta'_2=7489\}$, obtained from fitting \eqref{eq1} to the published experimental data. The criterion value of the reference design (i.e.\ the face-centred CCD) is $\phi=-52.7712$. 

Consider the full second-order polynomial model in terms of the scaled variables $x_1$, $x_2$ and $x_3$. This empirical model fits the data well, as shown in \citet[chapter 11]{bo}. For this reason, we also computed a LDOD for that model, but this LDOD $\mathit{X}_\mathrm{emp}$, shown in Table \ref{tab1.5}, does not improve much on the reference CCD for parameter estimation in the mechanistic Model \eqref{eq1} (the criterion value of $\mathit{X}_\mathrm{emp}$ is $-51.0181$). 

\begin{table}
\caption{\label{tab1.5} 24-Run LDOD for the Second Order Polynomial Model}
\centering
\begin{small}\begin{tabular}{c c c | c c c | c c c | c c c | c c c | c c c}
\hline
R&C&T&R&C&T&R&C&T&R&C&T&R&C&T&R&C&T\\[0.5ex]\hline
1.5&1&70&1.5&2&90&1.5&4&90&3&2&80&6&1&80&6&4&70\\
1.5&1&80&1.5&4&70&3   &1&70&3&4&80&6&1&90&6&4&70\\
1.5&1&90&1.5&4&70&3   &1&90&6&1&70&6&2&80&6&4&90\\
1.5&2&70&1.5&4&90&3   &2&70&6&1&70&6&2&90&6&4&90\\[0.01ex]\hline
\end{tabular}
\end{small}
\end{table}

Our interest is in the LDOD for Model \eqref{eq1}, the mechanistic approximation to the response surface, so we would expect to find a better design than those intended for fitting the polynomial model. To ensure an efficient design, we perform $\mathbb{\tau}=100$ independent tries with the critical value fixed at 1.0001. We use the factor levels in the reference CCD as the candidate coordinates, which are $\boldsymbol \Omega_1=\{1.5, 3, 6\}$, $\boldsymbol \Omega_2=\{1, 2, 4\}$ and $\boldsymbol \Omega_3=\{70, 80, 90\}$ for our implementation of the traditional CEA. The full candidate set $\boldsymbol \Omega$ for our implementation of the traditional PEA then consists of $3^3=27$ points. To speed up the iterative continuous optimisation, in our modified PEA and CEA, we instructed Matlab to use parallel computation. Using both the PEA and the CEA, we find the LDOD $\mathit{X}_\mathrm{d}$ in Table \ref{tab2}, the criterion value of which is $-49.7321$. With the PEA, the design in Table \ref{tab2} was found in three of the 100 tries. With the CEA, the design in Table \ref{tab2} was found only once. The best three designs are dissimilar and their criterion values are $-49.7321$, $-49.7452$ and $-49.7452$. 

\begin{table}
\caption{\label{tab2} 24-Run LDOD Obtained Using Discrete Optimisation for the Mechanistic Model}
\centering
\begin{small}\begin{tabular}{c c c | c c c | c c c | c c c | c c c | c c c}
\hline
R&C&T&R&C&T&R&C&T&R&C&T&R&C&T&R&C&T\\[0.5ex]\hline
1.5&1&70&1.5&1&90&1.5&4&80&1.5&4&90&6&1&80&6&4&70\\
1.5&1&80&1.5&4&70&1.5&4&90&3&1&70&6&1&90&6&4&70\\
1.5&1&90&1.5&4&70&1.5&4&90&3&1&70&6&1&90&6&4&80\\
1.5&1&90&1.5&4&70&1.5&4&90&6&1&80&6&4&70&6&4&80\\[0.01ex]\hline
\end{tabular}
\end{small}
\end{table}

To compare this with the discrete optimisation over $\boldsymbol \Omega$, we also do 100 independent tries and use the same critical value 1.0001. In each step of the iteration, the current point (or coordinate) also acts as the initial vector (or value) in optimising the criterion function. When the continuous PEA is used, it produces the design in Table \ref{tab3}(a), which has a local D-criterion value of $\phi=-49.5528$. The mean number of iterations is 4.4, so it does not take many steps to obtain a local maximum of $\phi$. When we use the continuous CEA, the $\mathit{X}_\mathrm{d}$ in Table \ref{tab3}(b) is found to be locally D-optimal. The maximal criterion value is $-49.5573$, which is worse than that of the design in Table \ref{tab3}(a). On average, the number of iterations is 5, so no clear difference between PEA and CEA is visible here. 

\begin{table}
\caption{\label{tab3} 24-Run LDODs with Continuous Optimisation}
\centering
\begin{small}\begin{tabular}{l c l | l c l ||| l c l | l c l}
\hline
\multicolumn{6}{c}{(a) PEA}&\multicolumn{6}{c}{(b) CEA}\\[0.5ex]\hline
~R&C&~T&~R&C&~T&~R&C&~T&~R&C&~T\\[0.5ex]\hline
1.5&1&70&2.9812&1&70&1.5&1&70&3.4081&1&70\\
1.5&1&90&2.9836&1&70&1.5&1&70&3.4102&1&70\\
1.5&1&90&2.9848&1&70&1.5&1&90&5.8728&4&70\\
1.5&1&90&5.8689&4&70&1.5&1&90&5.8747&4&70\\
1.5&4&73.9282&5.8691&4&70&1.5&1&90&5.8758&4&70\\
1.5&4&73.9316&5.8694&4&70&1.5&4&74.0128&5.8765&4&70\\
1.5&4&73.9444&6&1&85.0823&1.5&4&74.0131&6&1&84.5574\\
1.5&4&73.9458&6&1&85.1136&1.5&4&74.0250&6&1&84.7942\\
1.7667&4&90&6&1&85.1208&1.7304&4&90&6&1&84.8235\\
1.7667&4&90&6&1&85.1353&1.7304&4&90&6&1&84.9113\\
1.7668&4&90&6&4&78.4433&1.7305&4&90&6&4&80.1277\\
1.7668&4&90&6&4&78.4712&1.7306&4&90&6&4&80.1320\\[0.01ex]\hline
\end{tabular}
\end{small}
\end{table}

When using continuous optimisation, exact replicate runs are rarer than near-replicates, as we see in Table \ref{tab3}. This is because the design region $\mathcal{X}$ is continuous and implies an infinite number of feasible candidate runs. However, despite small differences in the later decimal places of the coordinates, it is clear what kind of coordinates we should use under the optimality criterion. Many of the runs are not far in distance from the candidates previously defined in the $3 \times 3 \times 3$ set $\boldsymbol \Omega$. Hence the designs in Table \ref{tab3} does not show much of an improvement over that in Table \ref{tab2}, which we found with the traditional discrete optimisation. With respect to Table \ref{tab3}(a) which consists of quasi-continuous coordinates (i.e.\ the real coordinate values are irrational numbers), the reference CCD is $\mathrm{exp}(-52.7712/6)/\mathrm{exp}(-49.5528/6) \approx 58.48\%$ efficient and $\mathit{X}_\mathrm{emp}$ is $78.33\%$ efficient. However, the relative efficiency is as high as $97.06\%$ for the design in Table \ref{tab2}, which is made up of candidate points from $\boldsymbol \Omega$. This shows that, even though the model is complex and nonlinear, the discrete optimisation can be effective provided a suitable candidate set is used.

\subsection{Example 2: Hybrid Nonlinear Model}
For the experiment in \citet{mo}, we assume Model \eqref{eq2}. To determine $\boldsymbol \theta^0$, we fit that model to the dataset obtained from the 18-run face-centred CCD (Table \ref{tab4}, where no valid response was obtained from the 16th run). The nonlinear least squares estimate is $\tilde{\boldsymbol \theta}=\{\tilde a_0,\tilde a_1 ,\tilde a_2,\tilde a_3,\tilde a_4,\tilde a_5\} \approx \{0.4340,1.3140,-0.1059,-0.8224,0.4105,-2.0633\}$, which we use as the prior $\boldsymbol{\theta}_0$ for choosing a new design. The local D-criterion function value $\phi$ of the reference CCD is $31.7538$ under $\boldsymbol{\theta}_0$. We assume the same design region for the new experiment, which is a cuboid $(S,E,P) \in \mathcal{X}=[2.5,7.5] \times [0.625,62.5] \times [200,400]$. Both enzyme concentration $E$ and pressure $P$ are scaled as in Model \eqref{eq2}. 

\begin{table}
\caption{\label{tab4} 18-Run Reference Face-Centred CCD and Response from \citet{mo}}
\centering
\begin{small}\begin{tabular}{c c c c | c c c c | c c c c }
\hline
S&E&P&$\xi$&S&E&P&$\xi$&S&E&P&$\xi$\\[0.5ex]\hline
5& 6.25& 300& 73.6&2.5&62.5&400&95.2&7.5&62.5&400&82.7\\
5& 6.25& 200& 81.6&7.5&6.25&300&77.3&2.5&6.25&300&90.0\\
5& 62.5& 300& 76.0&5~&6.25&400&69.0&2.5&0.625&400&55.2\\
5& 6.25& 300& 69.4&7.5&0.625&200&43.3&7.5&0.625&400&\_\\
5& 6.25& 300& 73.6&2.5&0.625&200&62.8&7.5&62.5&200&87.0\\
5& 0.625& 300& 50.5&5  &6.25&300&74.0&2.5&62.5&200&96.0\\[0.01ex]\hline
\end{tabular}
\end{small}
\end{table}

As an empirical approximation to the response surface, again, one can fit a full second-order linear model in terms of the scaled factors $x_{\Scale[0.5]{\mathrm{S}}}$, $x_{\Scale[0.5]{\mathrm{E}}}$ and $x_{\Scale[0.5]{\mathrm{P}}}$, where
$$x_{\Scale[0.5]{\mathrm{S}}}=\frac{S-5}{2.5} \in [-1,1].$$
It is straightforward to find the LDOD $\mathit{X}_\mathrm{emp}$ for the empirical model over discrete candidates. If we evaluate $\mathit{X}_\mathrm{emp}$ under Model \eqref{eq2}, the local D-criterion function value is $34.4783$, which is clearly better than that from the CCD. 

It can be shown that Model \eqref{eq2} gives a better approximation to the data than the polynomial model. The critical value 1.0001 is used in our search for an 18-run LDOD $\mathit{X}$ to maximise the local D-criterion function for Model \eqref{eq2}. For discrete optimisation with the PEA, the candidate set $\boldsymbol \Omega$ of $3^3$ points is based on the factor levels in the CCD in Table \ref{tab4}. The best three solutions obtained from 100 tries of that algorithm are similar. For each of them, the criterion function value is $\phi=38.8433$. Under the CEA, the largest three values obtained are $38.7313$, $38.6514$ and $38.6514$, so the PEA works better than the CEA in this case.

For the continuous optimisation, each exchange in the iterative procedure starts at multiple initial points, in addition to the current design point $\mathit{X}_i$. We use a coarse set of $2^3$ initial points $\{3.75,6.25\} \times \{1.9764,19.764\} \times \{250,350\}$ forming a cube within the design region $\mathcal{X}$ halfway between the centre and the edges of the region. To maximise the criterion function in $v=3$ factors, at each iteration, there are in total $(1+2^3)=9$ initial starts under the PEA and three for unidimensional maximisation under the CEA. The designs found using the continuous PEA and CEA are shown in Table \ref{tab5}(a) and (b) respectively. We can see little difference between these two solutions, which both have a D-criterion value of $41.2246$ to four decimal places. 

\begin{table}
\caption{\label{tab5} 18-Run LDOD with Continuous Optimisation}
\centering
\begin{footnotesize}\begin{tabular}{l l l | l l l ||| l l l | l l l}
\hline
\multicolumn{6}{c}{(a) PEA}&\multicolumn{6}{c}{(b) CEA}\\[0.5ex]\hline
~S&~~E&~~P&~S&~~E&~~P&~S&~~E&~~P&~S&~~E&~~P\\[0.5ex]\hline
2.5&2.7970&200&2.5&62.5&200&2.5&2.7970&200&2.5&62.5&200\\
2.5&2.7973&200&2.5&62.5&200&2.5&2.7970&200&2.5&62.5&200\\
2.5&2.7974&200&2.5&62.5&288.88&2.5&2.2978&200&2.5&62.5&288.88\\
2.5&19.275&400&2.5&62.5&288.88&2.5&19.276&400&2.5&62.5&288.88\\
2.5&19.280&400&2.5&62.5&400&2.5&19.279&400&2.5&62.5&400\\
2.5&20.647&200&2.5&62.5&400&2.5&20.641&200&2.5&62.5&400\\
2.5&20.648&200&3.0978&62.5&200&2.5&20.650&200&3.0978&62.5&200\\
2.5&23.248&288.37&3.1501&31.239&200&2.5&23.247&288.37&3.1502&31.243&200\\
2.5&23.249&288.37&3.1502&31.255&200&2.5&23.250&288.37&3.1502&31.246&200\\[0.01ex]\hline
\end{tabular}
\end{footnotesize}
\end{table}

Here the continuous optimisation results in a bigger improvement than in the previous example. Compared with the designs produced by continuous optimisation, shown in Table \ref{tab5}, the design obtained by discrete optimisation is $\mathrm{exp}(38.8433/6)/\mathrm{exp}(41.2246/6) \approx 67.24\%$ efficient. The design for the empirical model $\mathit{X}_\mathrm{emp}$ is $32.49\%$ efficient and the reference CCD is only $20.63\%$ efficient. This demonstrates that, if a nonlinear multifactor model is to be used, a design for this model must be sought, rather than relying on standard designs. It is also clear that the greatest benefit can only be realised by using a continuous optimisation algorithm.

\section{A Novel Multiphase Optimisation Algorithm}\label{sec:multiphase}
The continuous optimisation algorithm demands many tries to find a highly efficient design. This is due to the fact that random sampling to construct a starting design lacks effectiveness, because many initial designs cannot lead to D-optimal designs. In order to improve the effectiveness of the search, we introduce a refinement to the new continuous optimisation method. In most situations, there is insufficient information about the appropriate candidate points that would eventually compose the LDOD. However, it can be expected that the more efficient are the initial designs, the more efficient the final design will be under the D-criterion.

A new idea is to perform a swift discrete optimisation, over a small, coarse, candidate set, to convert each random initial design into an intermediate solution $\mathit{X}_\mathrm{d1}$, to obtain a reasonably efficient starting design for the continuous optmisation. To ensure a fast generation of the interim solutions in Phase 2 and to ensure that we get several different intermediate designs, we increase the critical value for this optimisation, e.g.\ to 1.1. This is \textbf{Phase 1} of the multiphase algorithm. Next, the distinct $\mathit{X}_\mathrm{d1}$ shall act as new starting designs for the continuous optimisation in \textbf{Phase 2}, the computation of which is more intensive. 

Practical limitations will influence how we can set the levels of the factors. To take account of them, in \textbf{Phase 3}, the closest distance of a factor indicates the minimum space between two feasible coordinates next to each other, the levels of which must be distinguishable to experimenters.  As we have set the $v$ closest distances, the continuous variable space $\mathcal{X}$ is redefined to be a discrete set of candidate levels. We then adjust the continuous design to use only levels from this discrete set. Our \textit{multiphase optimisation PEA} is as follows:

\begin{enumerate}
  \item[1.1] \textbf{Phase 1}: Compute the LDOD over discrete set $\boldsymbol \Omega$, using \textbf{STEPS 1-7} in the traditional PEA. 
  \item[1.2] Refer to the intermediate solutions obtained in \textbf{Phase 1} as $\mathit{X}_\mathrm{d1}$.
  \item[2.1] \textbf{Phase 2}: Compute the LDOD over $\mathcal{X}$, using either the continuous PEA or the contiuous CEA, for each $\mathit{X}_\mathrm{d1}$ acting as the starting design $\mathit{X}$.
  \item[2.2] Refer to the revised solutions as $\mathit{X}_\mathrm{d2}$, the most efficient of which is selected to be $\mathit{X}_\mathrm{d}$.
  \item[3.1] \textbf{Phase 3}: Find the LDOD for refinement. For $k=1,2,\ldots,v$, define the \textit{closest distance} for the $k$th factor. Points with all factors within the closest distance will be considered quasi-replicates. Group the quasi-replicates in $\mathit{X}_\mathrm{d}$, so that the $n$ runs of $\mathit{X}=\mathit{X}_\mathrm{d}$ are divided into $n^*$ homogeneous clusters. Sort the rows of $\mathit{X}$ with respect to the clusters and factor levels. Reset $k$ to 1.
  \item[3.2] Set both $i$ and $i^*$ to 1 as the starting values.
  \item[3.3] Pick out the $i^*$th cluster of $\mathit{X}$, corresponding to $n_{i^*}$ quasi-replicate points, which will be replaced by true replicates. Let the maximum value in the cluster of the $k$th factor be $X_{\mathrm{max},k}$ and the minimum be $X_{\mathrm{min},k}$.
  \item[3.4]  Create a provisional subset of candidate coordinates $\boldsymbol \Omega_k$, on the basis of the closest distance, the interval $[X_{\mathrm{min},k}, X_{\mathrm{max},k}]$, and $\mathcal{X}$. Search over $\boldsymbol \Omega_k$ and let $X_{\mathrm{new},k}$ be the candidate that maximises $\phi$. Use this substitute for each point of the $i^*$th cluster. Set $i$ to $i+n_{i^*}$. After substitution, there are $n_{i^*}$ identical values of the new coordinate. 
  \item[3.5] Unless $i=n+1$, set $i^*$ to $i^*+1$ and return to \textbf{STEP 3.3}.  
  \item[3.6] Unless $k=v$, set $k$ to $k+1$ and return to \textbf{STEP 3.2}.
  \item[3.7] Create a new candidate set $\boldsymbol \Omega$, consisting of the $n^*$ distinct design points of the current $\mathit{X}$. Perform a discrete optimisation over $\boldsymbol \Omega$, the final solution $\mathit{X}_\mathrm{d}^*$ of which is the LDOD.
\end{enumerate}

In this improved algorithm, \textbf{STEP 3.3} creates the variable space for factor $k$ in the $i^*$th cluster, and then \textbf{STEP 3.4} uses a variant of coordinate exchange to optimise the common factor level. The aim is to find the best level for the $k$th factor. Suppose the closest distance is 0.01 unit for the flow rate in the \citeauthor{bo} example. In Table \ref{tab3}b, the 13th and 14th runs are quasi-replicates. The corresponding two flow rates are within the interval $[3.4081, 3.4102] \subset [3.40,3.42]$. To select their factor levels as in \textbf{STEP 3.4}, the candidate subset should be $\boldsymbol \Omega_1=\{3.40,3.41,3.42\}$. In the final solution $\mathit{X}_\mathrm{d}^*$, $n^*$ is equal to the number of distinct design points and $n_{i^*}$ indicates the number of exact replicates at each point, for $i^*=1,2,\ldots,n^*$. 

The CEA can improve the continuous optimisation if the number of factors $v$ is large. Otherwise, the PEA is found to be more reliable in our examples. 

In the multiphase algorithm, the new continuous optimisation method is used to complement the basic discrete optimisation over an imperfect candidate set. \textbf{Phase 3} aims to convert the quasi-replicate runs of $\mathit{X}_\mathrm{d}$ into exact replicates. \citet{do} proposed another algorithm for adjusting an already efficient design, though this is not specifically to create replicates.

\section{Examples Revisited}\label{sec:moreresults}
\subsection{Example 1: Multifactor Mechanistic Model}
For the \citeauthor{bo} example, the new multiphase optimisation algorithm can start with the same $3 \times 3 \times 3$ candidate set $\boldsymbol \Omega$, based on the factor levels from the reference CCD. This candidate set is suitable for an initial discrete optimisation and also facilitates Phases 2 and 3. With 30 random tries and a critical value of 1.1 in Phase 1 of the multiphase algorithm, 30 distinct intermediate solutions $\mathit{X}_\mathrm{d1}$ are obtained using the traditional PEA. Next, in Phase 2, the continuous optimisation starts from each of the designs in that set using a critical value 1.0001 in Phase 2. Typical interim solutions after Phase 2 are shown in Table \ref{tab6}(a) and (b) using point exchange and coordinate exchange respectively. Using the continuous PEA, the mean number of iterations is 3.6 in Phase 2, while the largest three criterion values are $-49.5143$, $-49.5182$ and $-49.5182$ (the mean of the 30 is $-49.5677$). In comparison to the continuous optimisation results in Section \ref{sec:results}, where the mean criterion value of 100 tries is $-49.7496$ and the mean number of iterations is 4.4, there is a clear improvement in the quality of the designs and the number of iterations. Likewise, the continuous CEA takes 3.8 iterations on average in Phase 2 of the multiphase algorithm and the mean of the 30 criterion values is $-49.5687$ (the largest three values of which are $-49.5143$, $-49.5182$ and $-49.5282$). Thus the PEA should be preferred for Phase 2.

\begin{table}
\caption{\label{tab6} 24-Run LDOD for Model \eqref{eq1}: Interim Solution after Phase 2 of the Multiphase Method}
\centering
\begin{small}\begin{tabular}{l c l | l c l ||| l c l | l c l}
\hline
\multicolumn{6}{c}{(a) PEA}&\multicolumn{6}{c}{(b) CEA}\\[0.5ex]\hline
~R&C&~T&~R&C&~T&~R&C&~T&~R&C&~T\\[0.5ex]\hline
1.5&1&70&1.7348&4&90&1.5&1&70&1.7348&4&90\\
1.5&1&70&1.7348&4&90&1.5&1&70&1.7349&4&90\\
1.5&1&90&3.3188&1&70&1.5&1&90&3.3188&1&70\\
1.5&1&90&3.3190&1&70&1.5&1&90&3.3191&1&70\\
1.5&1&90&5.8118&4&70&1.5&1&90&5.8119&4&70\\
1.5&1&90&5.8120&4&70&1.5&1&90&5.8121&4&70\\
1.5&4&74.0877&5.8121&4&70&1.5&4&74.0840&5.8123&4&70\\
1.5&4&74.1142&6&1&84.8573&1.5&4&74.1156&6&1&84.8571\\
1.5&4&74.1295&6&1&84.8577&1.5&4&74.1303&6&1&84.8583\\
1.5&4&74.1343&6&1&84.8591&1.5&4&74.1329&6&1&84.8594\\
1.7348&4&90&6&4&79.3271&1.7348&4&90&6&4&79.3267\\
1.7348&4&90&6&4&79.3282&1.7348&4&90&6&4&79.3274\\[0.01ex]\hline
\end{tabular}
\end{small}
\end{table}

With the CEA, we also tested Brent's unidimensional method instead of the Nelder-Mead method. The CEA takes the maximum of 30 iterations in all but one case. The mean criterion value is $-50.0487$ and the largest three values are $-49.8285$, $-49.8633$ and $-49.8766$. Brent's method is faster for a single iteration, but it fails to deliver a stable and efficient solution after Phase 2. In our demonstration on a Windows desktop computer with an Intel Core i7 Processor, the total required computing time for 30 tries is 8 seconds with the PEA, 10 with the CEA using the Nelder-Mead algorithm, and 15 with the CEA using Brent's method. Thus, our improved algorithm takes just over half of the computing time of the best existing method, though in this case the difference is practically unimportant.

We can choose either interim solution $\mathit{X}_\mathrm{d}$ in Table \ref{tab6} for Phase 3. The closest distance is set to be 0.1 unit for $R$, 0.1 unit for $C$ and 1 unit for $T$. After the reallocation of experimental runs described in \textbf{STEP 3.7} (i.e.\ recalculating the optimal numbers of replicates for each distinct design point), the final solution $\mathit{X}_\mathrm{d}^*$ in Table \ref{tab7} is obtained. This design has only eight distinct design points, to estimate the eight parameters. The distinct design points are unequally replicated. We also note that only two levels of the catalyst concentration are used. The new maximal criterion value is $-49.5116$ which is better than all values obtained previously. More importantly, now we have fully achieved the objective to accurately identify the factor levels of every distinct design point (at desirable precision) as well as the corresponding number of replicates. A standard experimental design (e.g.\ CCD or Box-Behnken design) requires 13-15 distinct design points here, as illustrated in \textbf{APPENDIX \ref{sec:B}}, while a full factorial design requires at least $3^3=27$ distinct points. By using the multiphase optimisation algorithm, in addition to achieving higher efficiency and more reliable parameter estimation, we are able to simplify the initially constructed design, making it more practical and more efficient. In this example, there are only $n^*=8$ support points in $\mathit{X}_\mathrm{d}^*$, in contrast to the 11 distinct design points for the initial design in Table \ref{tab2} which was found using the discrete optimisation. 

\begin{table}
\caption{\label{tab7} 24-Run LDOD for Model \eqref{eq1} produced by the multiphase algorithm}
\centering
\begin{small}\begin{tabular}{c c c | c c c | c c c | c c c | c c c | c c c}
\hline
R&C&T&R&C&T&R&C&T&R&C&T&R&C&T&R&C&T\\[0.5ex]\hline
1.5&1&70&1.5&1&90&1.5&4&74&1.7&4&90&5.8&4&70&6&1&85\\
1.5&1&70&1.5&1&90&1.5&4&74&1.7&4&90&5.8&4&70&6&1&85\\
1.5&1&90&1.5&4&74&1.7&4&90&3.3&1&70&5.8&4&70&6&1&85\\
1.5&1&90&1.5&4&74&1.7&4&90&3.3&1&70&6&1&85&6&4&79\\[0.01ex]\hline
\end{tabular}
\end{small}
\end{table}

As a side note, even if we use Brent's method for CEA (i.e.\ unidimensional continuous optimisation), it is feasible to obtain Table \ref{tab7}. We can, for instance, increase the maximum number of iterations and reduce the critical value (currently at 1.0001) to allow for more optimisations and achieve a stable D-optimality. More effectively, we can carry out \textbf{STEP 3.7} of Phase 3 for every Phase 2 design $\mathit{X}_\mathrm{d2}$, as suggested by \citet{got} and implemented in JMP. Using JMP 13, with 100 tries using Brent's method, we found a design which is only slightly worse than that in Table \ref{tab7} and the same up to rounding of factor levels, though it takes longer to find.

\subsection{Example 2: Hybrid Nonlinear Model}
For the \citeauthor{mo} example, after the initial discrete optimisation (Phase 1), there are just four distinct designs $\mathit{X}_\mathrm{d1}$ that can be used as inputs for Phase 2, each of which is made up of 18 runs selected from the 24-run candidate set $\boldsymbol \Omega$. At the end of Phase 2, with the PEA, the best local D-criterion function values obtained are $41.2246$, $41.2052$, $41.2052$ and $41.1403$ whereas the mean number of iterations is 4. When we use the CEA, the criterion values of $\mathit{X}_\mathrm{d2}$ are $41.2246$, $41.1319$, $41.1301$ and $41.1301$, obtained after 4.5 iterations on average. We then set the closest distance to be 0.01 for $S$, 0.005 for $E$, and 0.1 for $P$. These are used in Phase 3 to obtain the design $\mathit{X}_\mathrm{d}^*$ in Table \ref{tab8}, the criterion value of which is also $41.2246$. This is similar to the D-criterion value of the design in Table \ref{tab5}, so the design does not improve the D-efficiency, but it simplifies the search without sacrificing D-efficiency. The new design has nine distinct design points, seven of which are replicated twice and one is replicated three times.

\begin{table}
\caption{\label{tab8} 18-Run LDOD for Model \eqref{eq2} produced by the multiphase algorithm}
\centering
\begin{small}\begin{tabular}{l l l | l l l | l l l}
\hline
S&E&P&S&E&P&S&E&P\\[0.5ex]\hline
2.5&2.795&200&2.5&20.645&200&2.5&62.5&288.9\\
2.5&2.795&200&2.5&23.25&288.4&2.5&62.5&400\\
2.5&2.795&200&2.5&23.25&288.4&2.5&62.5&400\\
2.5&19.275&400&2.5&62.5&200&3.1&62.5&200\\
2.5&19.275&400&2.5&62.5&200&3.15&31.25&200\\
2.5&20.645&200&2.5&62.5&288.9&3.15&31.25&200\\[0.01ex]\hline
\end{tabular}
\end{small}
\end{table}

\section{Discussion and Recommendations}\label{sec:discussion}
Empirical polynomial response surfaces are widely used and good designs for estimating them are readily available. However, as noted in \citet{bo}, experimenters might benefit from deriving a mechanistic model or an empirical nonlinear model, which can better approximate the response surface and facilitate the scientific interpretation of results. The optimal design of multifactor experiments for estimating nonlinear models has been rather neglected and the traditional design method via discrete optimisation is not always successful. To solve this problem, we have developed a multiphase optimisation algorithm. Searching for such an optimal design is no longer challenging with the multiphase optimisation method. When this algorithm is applied to both illustrative examples, we easily obtain a locally D-optimal design. These locally D-optimal designs have a simple structure that is different from those of standard experimental designs. It would therefore be interesting to apply the methodology in this paper to study different classes of nonlinear experiments. 

To promote such designs, there are two challenges in practice: (i)\ to find a suitable form of the nonlinear model; and (ii)\ to choose a set of realistic values as the parameter prior. On the first challenge, if we cannot assume a mechanistic model, often it is still possible to use a hybrid nonlinear model reflecting some characteristics of the underlying mechanism. This is demonstrated in Example 2. As to the second challenge, we should base the locally D-optimal design on the most realistic parameter values we are able to assume. Some reference data, either from the literature or earlier experiments, are helpful. Otherwise, values obtained from the literature on related systems can be useful, or in the absence of anything else experimenters have to use their knowledge, experience and intuition to suggest prior values. It is important to note that, even if these values are inaccurate, the design chosen will still be valid and usually quite efficient, even though it will not in general be optimal for the unknown true parameter values.

Our continuous optimisation method is robust in updating the factor levels in $\mathit{X}$, as we demonstrate in both examples. It takes the whole design space $\mathcal{X}$ into account, rather than a finite candidate set $\boldsymbol \Omega$. Hence, we avoid the risk of using an improperly specified candidate set. Further, the multiphase optimisation algorithm circumvents ineffective tries and iterations, in order to find a locally D-optimal design with less computational effort. 

The most important message from our work is that, if a multifactor nonlinear model is to be used to gain scientific insight from experimental data analysis, it is important to tailor the experimental design to that model. We have provided the tools needed to do this effectively and we recommend their use to experimenters in chemical engineering, biochemistry, and related areas. The enormous potential to harvest data which are informative about plausible mechanisms, rather than just giving rough predictions more than compensates for the extra efforts taken to acquire designs. 

\section{Acknowledgements}
We thank the referees and Associate Editor for extensive comments which have helped us to improve and clarify the paper. The first author acknowledges a Vice-Chancellor's Scholarship from the University of Southampton and financial support from the Faculty of Applied Economics of the University of Antwerp. The third author has received funding from the Universidad Carlos III de Madrid, the European Union's Seventh Framework Programme for research, technological development and demonstration under grant agreement no.\ 600371, el Ministerio de Economía y Competitividad (COFUND2013-40258), el Ministerio de Educaci\'{o}n, cultura y Deporte (CEI-15-17) and Banco Santander.

\bibliography{9months}

\appendix
\section{Derivation of the Multifactor Model for Example 2}\label{sec:A}
For the \citet{mo} experiment, we are interested in a model of the substrate conversion rate $\xi$, in terms of the initial substrate concentration $S$. The reaction mechanism is unknown but it involves a mixture of two reagents: the substrate (dextran) and an activator (i.e. enzyme endodextranase). It is therefore reasonable to assume the first step of mechanism to be reversible and bimolecular: the small molecules of dextran will attach to the active sites of endodextranase molecules. A bond like this could form an intermediate complex in an unstable state, which is then converted to the final reaction product.

Under an extreme scenario when even the maximal substrate concentration $S=7.5$ (\% weight/volume) is far too low (relative to the fixed concentration of endodextranase), the reaction rate will be high and stable over time. The conversion is then close to 100\% at the measurement time (after the reaction almost ceases), when the instantaneous substrate concentration is near zero and treated as independent of the initial concentration. The expected conversion rate is $\mathbb{E}(\xi) \rightarrow 100(S-a)/S$, where $a$ is a nonnegative constant near zero. In contrast, if even the minimal substrate concentration level 2.5 is too high for the concentration of endodextranase, the reaction will be slow and its rate can decrease quickly over time. The concentration of the converted substrate is almost independent of the initial substrate concentration, so that $\mathbb{E}(\xi) \rightarrow 100b/S$, where $b$ is a nonnegative constant. 

A tradeoff must be found between these two impractical scenarios, where we assume an ideal solution (or ideal mixture). When the substrate concentration increases, $\xi$ shall decrease. It is also more realistic to consider the function of $\xi$ to be concave rather than convex. As an approximation to the observed response surface in \citet{mo}, we consider a nonlinear function in parameters $\gamma_1\geq 0$ and $\gamma_2 \in (-2.5,0)$:
$$\mathbb{E}(\xi_i)=\frac{\gamma_1S_i}{\gamma_2+S_i}\text{, for}~i=1,2,\ldots,n.$$
To use a transformed dependent variable, we derive the statistical model
\begin{align}
\frac{\xi_i}{100-\xi_i}=\frac{\gamma_1'S_i}{\gamma_2'+S_i}+\varepsilon_i'\text{, for}~ i=1,2,\ldots,n,
\label{eq4}
\end{align}
which incorporates an error $\varepsilon_i'$ and two constants $\gamma_1' \ge 0$ and $\gamma_2' \in (-2.5,0)$. When the substrate concentration is fixed, we use a second-order linear function to explain the transformed conversion rate in the scaled variables $x_{\Scale[0.5]{\mathrm{E}}}$ and $x_{\Scale[0.5]{\mathrm{P}}}$. With interaction $x_{\Scale[0.5]{\mathrm{E}}}x_{\Scale[0.5]{\mathrm{P}}}$ excluded because of insignificance in this case, the final model is
\begin{align}
\frac{\xi_i}{100-\xi_i}=\mathrm{exp}(a_0'+a_1'x_{\Scale[0.5]{\mathrm{E}},i}+a_2'x_{\Scale[0.5]{\mathrm{P}},i}+a_3'x_{\Scale[0.5]{\mathrm{E}},i}^2+a_4'x_{\Scale[0.5]{\mathrm{P}},i}^2)+\varepsilon_i'',
\label{eq5}
\end{align}
where $\varepsilon_i''$ is the error and $a_0'$, $a_1'$, $a_2'$, $a_3'$ and $a_4'$ are unknown parameters. We combine \eqref{eq4} and \eqref{eq5} to write the nonlinear multifactor model as
$$\frac{\xi_i}{100-\xi_i}=\frac{\mathrm{exp}(a_0+a_1x_{\Scale[0.5]{\mathrm{E}},i}+a_2x_{\Scale[0.5]{\mathrm{P}},i}+a_3x_{\Scale[0.5]{\mathrm{E}},i}^2+a_4x_{\Scale[0.5]{\mathrm{P}},i}^2)S_i}{a_5+S_i}+\varepsilon_i\text{, for}~ i=1,2,\ldots,n,$$
where $\varepsilon_i$ is the error and $a_0$, $a_1$, $a_2$, $a_3$, $a_4$ and $a_5$ are the parameters. This model fits the reference data well, so we can use it to approximate the unknown response surface.

\section{Standard Designs for Example 1}\label{sec:B}
\setcounter{table}{0}
\setcounter{figure}{0}
\renewcommand{\thetable}{B\arabic{table}}
\renewcommand{\thefigure}{B\arabic{figure}}

Tables B1-B3 below show three standard experimental designs we may construct for Example 1, each of which consists of 24 independent runs. For each design, we define four exact replicate runs at the centre $(3, 2, 80)$ of the variable space $\mathcal{X}$. These designs are used in this paper as benchmarks in evaluating the relative efficiency of the LDODs found using different algorithms in Sections \ref{sec:results} and \ref{sec:moreresults}. Particularly, the face-centred CCD is the reference design we choose and study in Section \ref{sec:results}, the local D-criterion value of which is $-52.7712$. Note that the spherical CCD (Table B2) is even less efficient in this case, with a criterion value of $-54.6880$. This is because of the suboptimal factor levels it uses. The modified Box-Behnken design (BBD) (Table B3), which has several points replicated more than a standard BBD, has a larger criterion value at $-51.5174$, but is still less efficient than any optimal designs we find. Overall, while it is easy to construct these standard designs when exactly three factors are under control, they are not very useful for estimating the parameters of the nonlinear model.

\begin{table}
\caption{24-Run Reference Face-Centred CCD}\centering
\begin{small}\begin{tabular}{l l l | l l l | l l l | l l l}\hline
R&C&T&R&C&T&R&C&T&R&C&T\\[0.5ex]\hline
1.5&1&70&3&1&80&3&2&80&6&1&70\\
1.5&1&90&3&1&80&3&2&80&6&1&90\\
1.5&2&80&3&2&70&3&2&90&6&2&80\\
1.5&2&80&3&2&70&3&2&90&6&2&80\\
1.5&4&70&3&2&80&3&4&80&6&4&70\\
1.5&4&90&3&2&80&3&4&80&6&4&90\\[0.01ex]\hline
\end{tabular}\end{small}\label{B1}\end{table} 

\begin{table}
\caption{24-Run Spherical CCD}\centering
\begin{small}\begin{tabular}{l l l | l l l | l l l | l l l}\hline
R&C&T&R&C&T&R&C&T&R&C&T\\[0.5ex]\hline
1.5&2&80&3&1&80&3&2&80&4.8976&1.2251&72.9289\\
1.5&2&80&3&1&80&3&2&80&4.8976&1.2251&87.0711\\
1.8376&1.2251&72.9289&3&2&70&3&2&90&4.8976&3.2651&72.9289\\
1.8376&1.2251&87.0711&3&2&70&3&2&90&4.8976&3.2651&87.0711\\
1.8376&3.2651&72.9289&3&2&80&3&4&80&6&2&80\\
1.8376&3.2651&87.0711&3&2&80&3&4&80&6&2&80\\[0.01ex]\hline
\end{tabular}\end{small}\label{B2}\end{table} 

\begin{table}
\caption{24-Run Box-Behnken Design}\centering
\begin{small}\begin{tabular}{l l l | l l l | l l l | l l l}\hline
R&C&T&R&C&T&R&C&T&R&C&T\\[0.5ex]\hline
1.5&1&80&3&1&70&3&2&80&6&1&80\\
1.5&1&80&3&1&70&3&2&80&6&1&80\\
1.5&2&70&3&1&90&3&4&70&6&2&70\\
1.5&2&90&3&1&90&3&4&70&6&2&90\\
1.5&4&80&3&2&80&3&4&90&6&4&80\\
1.5&4&80&3&2&80&3&4&90&6&4&80\\[0.01ex]\hline
\end{tabular}\end{small}\label{B3}\end{table} 
\end{document}